\pdfoutput=1
\documentclass[a4paper]{article}
\usepackage{lipsum}
\usepackage[utf8]{inputenc}
\usepackage[left=2cm, top=2cm, right=2cm, bottom=2cm]{geometry}
\usepackage{amsmath,amssymb,amsfonts}
\usepackage{mathrsfs}
\usepackage{url}
\usepackage{optidef}
\usepackage[font=footnotesize]{subcaption}
\usepackage{algcompatible}
\usepackage{algorithm}
\usepackage{cite}
\usepackage[pdftex]{graphicx}
\usepackage{float}
\usepackage{pgfplots}
\pgfplotsset{compat=1.18}
\usepgfplotslibrary{groupplots}
\usepackage{pgfplotstable} 
\usepackage{siunitx}
\usepackage{bm}
\usepackage{colortbl}
\usepackage{xcolor}
\usepackage{booktabs}
\usepackage{titlesec}
\titleformat{\section}[block]
  {\normalfont\Large\bfseries}
  {\thesection}{1em}{}
\titleformat{\subsection}[block]
  {\normalfont\large\bfseries}
  {\thesubsection}{1em}{}
\titleformat{\subsubsection}[block]
  {\normalfont\normalsize\bfseries}
  {\thesubsubsection}{1em}{}

\usepackage[hidelinks]{hyperref}
\usepackage{dsfont}
\usepackage{wasysym}
\usepackage{array}
\usepackage{multirow}
\usepackage{easyReview}
\usepackage{comment}
\usepackage{tikz}
\usepackage{enumitem}
\usepackage{amsthm}

\theoremstyle{plain}

\usepackage{txfonts}
\usepackage{setspace}
\begin{document}

\title{\bfseries Mixed-Integer Second-Order Cone Programming for Multi-period Scheduling of Flexible AC Transmission System Devices}
\date{}
\maketitle

\vspace{-1em}
\begin{center}
Mohamad Charara\textsuperscript{1*}, Martin De Montigny\textsuperscript{2}, Nivine Abou Daher\textsuperscript{2}, \\
Hanane Dagdougui\textsuperscript{1}, Antoine Lesage-Landry\textsuperscript{1}

\vspace{1em}

\textsuperscript{1}Polytechnique Montréal, GERAD \& MILA, Canada \\
\textsuperscript{2}Hydro-Québec, Canada
\end{center}

\vspace{1em}

\section*{\normalsize\textcolor{black}{SUMMARY}}
With the increasing energy demand and the growing integration of renewable sources of energy, power systems face operational challenges such as overloads, losses, and stability concerns, particularly as networks operate near their capacity limits. Flexible alternating current transmission system (FACTS) devices are essential to ensure reliable grid operations and enable the efficient integration of renewable energy. This work introduces a mixed-integer second-order cone programming (MISOCP) model for the multi-period scheduling of key FACTS devices in electric transmission systems. The proposed model integrates four key control mechanisms: (i) on-load tap changers (OLTCs) for voltage regulation via discrete taps; (ii)  static synchronous compensators (STATCOMs) and (iii) shunt reactors for reactive power compensation; and (iv) thyristor-controlled series capacitors (TCSCs) for adjustable impedance and flow control. The objective is to minimize active power losses using a limited number of control actions while meeting physical and operational constraints at all times throughout the defined time horizon. To ensure tractability, the model employs a second-order cone relaxation of the power flow. Device-specific constraints are handled via binary expansion and linearization: OLTCs and shunt reactors are modelled with discrete variables, STATCOMs through reactive power bounds, and TCSCs using a reformulation-linearization technique (RLT). A multi-period formulation captures the sequential nature of decision making, ensuring consistency across time steps. The model is evaluated on the IEEE 9-bus, 30-bus, and RTS96 test systems, demonstrating its ability to reduce losses, with potential applicability to larger-scale grids.
\vspace{-0.3cm}
\section*{\normalsize\textcolor{black}{KEYWORDS}} 
Conic Relaxation, FACTS, Mixed-integer Convex Optimization, Optimization, Power Loss, Shunt Reactor, STATCOM, Tap Changer, TCSC, Voltage Control.
\section{Introduction}
Modern power systems face a growing complexity due to the increasing electricity demand and the integration of renewable source of energy causing a need for operational flexibility~\cite{kundur1994power, hatziargyriou2020advanced}. Flexible alternative current transmission system (FACTS) devices—such as on-load tap changers (OLTCs), shunt reactors and capacitors, static synchronous compensators (STATCOMs), and thyristor-controlled series capacitors (TCSCs)---are usually employed to regulate voltages and optimize power flows~\cite{hingorani2000facts, zhang2012facts}. Coordinating these heterogeneous devices over time is challenging due to the nonlinear and nonconvex nature of power flows, discrete control settings, and temporal dependencies. Existing models often fail to capture switching constraints, inter-device interactions, or power flow feasibility over time~\cite{wood2013power, zhu2015optimization}. To address these challenges, we propose a joint optimization model formulated as a mixed-integer second-order cone program (MISOCP). The proposed framework models discrete actions via a binary expansion for OLTCs and shunts, continuous variables along with switching control for STATCOMs, and captures TCSCs effects using reformulation-linearization techniques (RLT). The second-order cone relaxation of the power flow ensures tractability~\cite{jabr2006radial}, enabling multi-device loss minimization across time under a maximum control action constraint. Experiments on the IEEE 9-bus, 30-bus, and RTS96 test systems illustrates the performance over static or device-specific formulations.

\textit{Related Work.} 
 Aigner et al.~\cite{aigner2023globalopf} present an optimization framework for alternative current optimal power flow (AC-OPF) problems with discrete decisions, employing piecewise linear relaxations of the power flow resulting in a mixed-integer linear programs (MILP) which is iteratively refined. The study does not focus on FACTS devices or multi-period scheduling. Reference\cite{kayacik2021misocp} develops a MISOCP formulation for the optimal reactive power dispatch (ORPD) problem, integrating discrete OLTC tap ratios and shunt capacitor switching via binary expansion and relaxation-tightening techniques. Their analysis is restricted to single-period scenarios and does not consider of series compensation devices. The authors of~\cite{han2023bumpless} propose a two-stage convex optimization approach for coordinated transmission switching and FACTS control, integrating line switching, STATCOMs, and TCSCs in a transient-free transition model. While effective for ensuring static and dynamic performance during transitions, the method does not address long-term planning or include OLTC controls.
 Ibrahim et al.~\cite{ibrahim2022alternating} propose a multi-period ORPD formulation with voltage security constraints to enhance reactive reserve margins in transmission networks. The decomposition strategy alternates between local optimization with an MILP-based consensus coordination to jointly manage switching shunts, voltage magnitude setpoints, and OLTC decisions. However, the model lacks TCSC integration and relies on a linearized power flow model rather than a second-order cone programming technique. Dey et al.~\cite{dey2022power} and Mumtahina et al.~\cite{mumtahina2023literature} employ particle swarm optimization methods for STATCOM placement and sizing, respectively. While capable of handling nonlinearities, these heuristic approaches lack optimality guarantees and do not support time-coupled decisions or constraints. Reference~\cite{aljumah2024tcsc} investigates TCSC optimal placement using gradient-based optimization algorithm, aiming to mitigate line overloads through series compensation. Despite its effectiveness in reducing losses, the model lacks integration with other FACTS devices and does not consider operational coupling over time. Tarafdar Hagh et al.~\cite{tarafdar2025factsreview} provide a comprehensive review of FACTS devices, their modelling, and their applications in optimization frameworks. The authors summarize advances in MILP, MISOCP, and metaheuristic approaches and identify the benefits of coordinated multi-device control. However, no unified time-coupled MISOCP formulation is presented.

To the best of our knowledge, there is currently no work that formulates a joint, time-coupled mixed-integer convex framework capable of coordinating OLTCs, STATCOMs, TCSCs, and shunt devices subject to operational and control constraints. In this work, we propose a unified MISOCP framework for coordinated FACTS device scheduling. Our main contributions are as follows:

\begin{itemize}
    \item We develop a multi-period mixed-integer second-order cone programming (MISOCP) model for the joint optimization of OLTCs, STATCOMs, TCSCs, and shunt reactors, subject to switching and operational constraints.
    \item We model device behaviours using binary expansion for discrete tap and shunt selection, linear bounds for STATCOM injections, and reformulation-linearization (RLT) techniques for the convex modelling of TCSC effects~\cite{bingane2019tight}.
    \item We numerically test our model on the IEEE 9-bus, 30-bus, and RTS96 systems and illustrate its ability to reduce losses more effectively than static or single-device baselines.
\end{itemize}

\section{Methodology and Problem Formulation}
This section presents the methodological framework and mathematical formulation of the proposed MISOCP model for scheduling the control actions of key FACTS devices in power transmission systems. The objective is to improve operational efficiency by minimizing active power losses while ensuring compliance with network constraints and device capabilities. The formulation integrates device-specific models for OLTCs, STATCOMs, shunt reactors, and TCSCs within a multi-period optimization framework, and leverages convex relaxations and binary encodings to ensure tractability and solution accuracy.

\subsection{Notation}
For all the following formulation, let $\mathcal{T} = \{1, \dots, T\}$ be the time horizon, where $T \in \mathbb{N}$.  
Let $\mathcal{N} \subseteq \mathbb{N}$ be the set of network nodes.  
Let $\mathcal{N}^{\text{OLTC}} \subseteq \mathcal{N}$ be the subset of nodes where OLTCs are placed.  
Let $\mathcal{L} \subset \mathcal{N} \times \mathcal{N}$ be the set of transmission lines connecting node pairs.  
Let $\mathcal{L}^{\text{TCSC}} \subseteq \mathcal{L}$ be the subset of lines where TCSCs are placed.  
Let $\mathcal{M} = \{\text{shunt}, \text{stat}, \text{oltc}, \text{tcsc}\}$ be the set of device types. Let $\mathcal{D} = \{(d, i, m, \ell) \mid d \in \{1, \dots, D\},\, i \in \mathcal{N},\, m \in \mathcal{M},\, \ell \in \mathcal{L} \}$ be the set of devices, where each device is identified by an index $d$, a bus location $i$, a type $m$, and an associated line $\ell$ (if applicable, e.g., for TCSC only), and where $D \in \mathbb{N}$ is the number of devices. In the sequel, we use the notation $\text{dev} = (d, i, m, \ell) \in \mathcal{D}$ to denote the quadruplet referring to the device dev. For example, $\text{shunt} \in \mathcal{D}$ denotes a device with identification number $d \in \{1,2,\dots,D\}$ located at node $i \in \mathcal{N}^{\text{shunt}}$ and of type $m = \text{shunt}$. Let $\mathcal{K}_{\text{tap}} = \{1, \dots, K\}$ be the index set for the binary variables representing the discrete tap positions of an OLTC, where $K \in \mathbb{N}$ is the number of available tap steps.  
Let $\mathcal{K}_{\text{shunt}} = \{1, \dots, N\}$ be the index set for the binary variables representing the discrete reactive power blocks of the shunt device, where $N \in \mathbb{N}$ is the number of available blocks.

\subsection{Devices modelling constraints}
To reflect the operational behaviour of various FACTS devices, we introduce constraint formulations that account for their discrete settings, switching logic, and physical limits in a mixed-integer convex manner.

\subsubsection{Shunt reactors}
Reactive shunts and reactive shunt compensators have constraint limits on their injections. Let $q_{\text{shunt}}^t \in \mathbb{R_+}$ be the reactive power injected by a shunt $ \in \mathcal{D}$ at time $t$, and let $\underline{q}_{\text{shunt}}, \overline{q}_{\text{shunt}} \in \mathbb{R_+}$ be its respective bounds. These variables are constrained as follows \cite{sulaiman2022optimal, khan2021novel}:

\begin{equation}
    \underline{q}_{\text{shunt}} \leq q_{\text{shunt}}^t \leq \overline{q}_{\text{shunt}} \ .
    \label{eq:qshunt_bounds}
\end{equation}
We assume that only fixed capacitive reactive power blocks can be aggregated and injected as needed. To represent this behaviour, we employ a binary expansion. Let $\bm{\alpha}_{\text{shunt}}^t \in \{0,1\}^N$ be the binary vector indicating the reactive power blocks selected of a $\text{shunt} \in \mathcal{D}$ at time $t$, where $\bm{\alpha}_{\text{shunt}}^t$ is an one-hot vector and $N$ is the number of blocks available in the list of indexes $\mathcal{K}_{\text{shunt}}$. Let $\mathbf{q}_{\text{shunt}} \in \mathbb{R}^{N}_+$ be the parameter vector containing the discrete reactive power possible values of a $\text{shunt} \in \mathcal{D}$. Let $s_{\text{shunt}}^t \in \{0,1\}$ be the binary status of a shunt $\in \mathcal{D}$ at time $t$, where $s_{\text{shunt}}^t = 1$ if the shunt is \textsc{ON}, and $0$ otherwise. Let $M \in \mathbb{R}_+$ be a Big-M constant used to deactivate constraints when the device is \textsc{OFF}. The constraints of reactive shunts are: 
\begin{alignat}{2}
    s_{\text{shunt}}^t &\leq \sum_{n \in \mathcal{K}_{\text{shunt}}} \alpha_{\text{shunt},n}^t 
    \leq N
    && \label{eq:shunt_one_hot} \\
    \sum_{n \in \mathcal{K}_{\text{shunt}}} \alpha_{\text{shunt},n}^t q_{\text{shunt},n} - (1 - s_{\text{shunt}}^t) M 
    &\leq q_{\text{shunt}}^t 
    \leq \sum_{n \in \mathcal{K}_{\text{shunt}}} \alpha_{\text{shunt},n}^t q_{\text{shunt},n} + (1 - s_{\text{shunt}}^t) M
    \label{eq:shunt_upper_bigM} \\
    0 
    &\leq q_{\text{shunt}}^t 
    \leq M s_{\text{shunt}}^t \ .
    \label{eq:shunt_Mcap}
\end{alignat}
The reactive power injection at time $t$ in \eqref{eq:shunt_upper_bigM} is determined by the sum of the selected discrete reactive power blocks, each weighted by its corresponding value. This summation captures the inner product between the selection vector $\bm{\alpha}_{\text{shunt}}^t$ and the parameter vector $\mathbf{q}_{\text{shunt}}$. If the shunt is inactive at time $t$, i.e., $s_{\text{shunt}}^t = 0$, the reactive power injection is forced to be zero using~\eqref{eq:shunt_Mcap}. Then,~\eqref{eq:shunt_one_hot} relaxes the one-hot condition by allowing the selection of up to $N$ blocks, while~\eqref{eq:shunt_upper_bigM} and~\eqref{eq:shunt_Mcap} replace~\eqref{eq:qshunt_bounds} to discretize the shunt behaviour and incorporate switching control.

\subsubsection{STATCOMs}
STATCOMs are devices that can inject or absorb reactive power. Their behaviour is usually nonlinear and depends on the voltage and phase. Let $q_{\text{stat}}^t \in \mathbb{R}$, $|v_{\text{stat}}^t| \in \mathbb{R}_{+}$, and $\phi_{\text{stat}}^t \in \mathbb{R}_{+}$ be the reactive power injection, voltage magnitude, and phase angle, respectively, of $\text{stat} \in \mathcal{D}$ at time $t$. Let $\underline{q}_{\text{stat}}, \overline{q}_{\text{stat}} \in \mathbb{R}$, $\underline{v}_{\text{stat}}, \overline{v}_{\text{stat}} \in \mathbb{R}_{+}$, and $\underline{\phi}_{\text{stat}}, \overline{\phi}_{\text{stat}} \in \mathbb{R}_{+}$ be their respective minimum and maximum bounds. The operational range is modelled by \cite{dey2022power, mumtahina2023literature}:
\begin{alignat}{2}
    \underline{q}_{\text{stat}} \; &\leq q_{\text{stat}}^t &&\leq \overline{q}_{\text{stat}} \label{eq:q_stat_limits} \\
    \underline{v}_{\text{stat}} \; &\leq |v_{\text{stat}}^t| &&\leq \overline{v}_{\text{stat}} \label{eq:v_stat_limits} \\
    \underline{\phi}_{\text{stat}} \; &\leq \phi_{\text{stat}}^t &&\leq \overline{\phi}_{\text{stat}} \ . \label{eq:phi_stat_limits}
\end{alignat}
For convexity purpose, we assume that a STATCOM operates only by injecting or absorbing reactive power directly. We further assume that the operating range of $|v_{\text{stat}}^t|$ matches the voltage magnitude limits at the STATCOM bus. Therefore, \eqref{eq:v_stat_limits} and \eqref{eq:phi_stat_limits} can be removed. We introduce an additional level of control to activate the device only when it is optimal. Let $s_{\text{stat}}^t \in \{0,1\}$ be the binary status of a $\text{stat} \in \mathcal{D}$ at time $t$, where $s_{\text{stat}}^t = 1$ if the device is \textsc{ON}, and $0$ otherwise. We substitute~\eqref{eq:q_stat_limits} with the following constraints:
\begin{align}
    |q_{\text{stat}}^t| &\leq \overline{q}_{\text{stat}} s_{\text{stat}}^t \ . \label{eq:stat_q_abs_bigM}
\end{align}
We retain \eqref{eq:stat_q_abs_bigM} to enforce operational bounds and reactive power injection by STATCOMs.

\subsubsection{OLTCs}
OLTCs regularize the voltage by controlling the transformer tap settings. Their nonconvex behaviour is described in \cite{iria2019optimal}. Let $|v_{\text{tap}}^t| \in \mathbb{R}_{+}$ be the regulated-side voltage magnitude of the transformer, $|\tilde{v}_{\text{tap}}^t| \in \mathbb{R}_{+}$ be the base-side voltage magnitude of the transformer, and $\varphi_{\text{tap}}^t \in \mathbb{R}_{+}$ be the turn ratio at time $t$ for a $\text{tap} \in \mathcal{D}$. Let $u_{\text{tap}}^t \in \mathcal{K}_{\text{tap}}$ be the discrete tap position index at time $t$. Let $\underline{\varphi}_{\text{tap}} \in \mathbb{R}_{+}$ be the minimum turn ratio and $\Delta \varphi_{\text{tap}} \in \mathbb{R}_{+}$ the incremental change per step. The OLTC operation is described by:
\begin{align}
    |v_{\text{tap}}^t| &= |\tilde{v}_{\text{tap}}^t| \varphi_{\text{tap}}^t \label{eq:tap_voltage_relation} \\
    \varphi_{\text{tap}}^t &= \underline{\varphi}_{\text{tap}} + \Delta \varphi_{\text{tap}} u_{\text{tap}}^t \label{eq:tap_ratio_definition} \\
    0                      &\leq u_{\text{tap}}^t \leq K \label{eq:tap_position_bound} \\
    \underline{v}_{\text{tap}} &\leq |\tilde{v}_{\text{tap}}^t| \leq \overline{v}_{\text{tap}} \ . \label{eq:tap_base_voltage_limits}
\end{align}
According to \cite{iria2019optimal},~\eqref{eq:tap_voltage_relation} – \eqref{eq:tap_base_voltage_limits} can be linearized using a binary expansion and the Big-M method. Let $\bm{\alpha}_{\text{tap}}^t \in \{0,1\}^K$ be the binary vector indicating the tap position of a $\text{tap} \in \mathcal{D}$ at time $t$, where the vector is one-hot encoded and $K$ is the number of available discrete steps. Let $\bm{\Delta\alpha}_{\text{tap}} \in \mathbb{R}_{+}^K$ be the vector of incremental turn ratio steps associated with each tap. Let $\tilde{U}_{\text{tap}}^t \in \mathbb{R}_{+}$ be the squared base-side voltage magnitude and $U_{\text{tap}}^t \in \mathbb{R}_{+}$ be the squared regulated voltage magnitude of the OLTC at time $t$. The selected tap ratio is computed by linearly combining the active tap with its corresponding step size, while enforcing a one-hot condition on the tap vector. The resulting squared voltage magnitude formulation is given by:
\begin{align}
    \varphi_{\text{tap}}^t &= \underline{\varphi}_{\text{tap}} + \sum_{n \in \mathcal{K}_{\text{tap}}} \alpha_{\text{tap},n}^t \Delta \alpha_{\text{tap},n} \label{eq:tap_phi_binary} \\
    \sum_{n \in \mathcal{K}_{\text{tap}}}\alpha_{\text{tap},n}^t &= 1\label{eq:tap_binary_one_hot} \\
    U_{\text{tap}}^t &= \tilde{U}_{\text{tap}}^t (\varphi_{\text{tap}}^t)^2 \label{eq:tap_voltage_squared} \\
    \underline{v}_{\text{tap}}^2 &\leq \tilde{U}_{\text{tap}}^t \leq \overline{v}_{\text{tap}}^2 \ . \label{eq:tap_squared_voltage_bounds}
\end{align}
Because~\eqref{eq:tap_voltage_squared} is still nonconvex, \cite{iria2019optimal} modifies \eqref{eq:tap_phi_binary} and \eqref{eq:tap_voltage_squared} using a bilinearization method. In our model, we convexify and linearize it by listing the limited and discrete values $(\varphi_{\text{tap}}^t)^2$ can take. Let $\bm{\delta}_{\text{tap}} \in \mathbb{R}_{+}^K$ be a parameter vector whose elements are given by $\delta_{\text{tap},n} = \left( \underline{\varphi}_{\text{tap}} + \Delta \alpha_{\text{tap},n} \right)^2$
 for all $n \in \mathcal{K}_{\text{tap}}$. Considering that $\bm{\alpha}_{\text{tap}}^t$ is a binary one-hot vector, it follows that $(\alpha_{\text{tap},n}^t)^2 = \alpha_{\text{tap},n}^t$, and the squared turn ratio $(\varphi_{\text{tap}}^t)^2$ can be expressed as:
\begin{equation}
    (\varphi_{\text{tap}}^t)^2 = \sum_{n \in \mathcal{K}_{\text{tap}}} \alpha_{\text{tap},n}^t \delta_{\text{tap},n} \ .
    \label{eq:phi_squared_linear}
\end{equation}
Using~\eqref{eq:phi_squared_linear}, we convexify~\eqref{eq:tap_voltage_squared}. Let $\bm{\gamma_{\text{tap}}}^t \in \mathbb{R}_{+}^K$ be an auxiliary variable and let $\gamma_{\text{tap},n}^t$ denote its $n^\text{th}$ component. We obtain the following constraints:

\begin{align}
    0 &\leq \gamma_{\text{tap},n}^t \leq M \alpha_{\text{tap},n}^t \label{eq:gamma_simple_bounds} \\
    \tilde{U}_{\text{tap}}^t \delta_{\text{tap},n} - M (1 - \alpha_{\text{tap},n}^t)
    &\leq \gamma_{\text{tap},n}^t
    \leq \tilde{U}_{\text{tap}}^t \delta_{\text{tap},n} + M (1 - \alpha_{\text{tap},n}^t) \label{eq:gamma_bigM_bounds} \\
    U_{\text{tap}}^t &= \sum_{n \in \mathcal{K}_{\text{tap}}} \gamma_{\text{tap},n}^t \ . \label{eq:gamma_sum}
\end{align}
Finally, we model the OLTC devices in a mixed-integer convex manner using ~\eqref{eq:tap_binary_one_hot}, \eqref{eq:tap_squared_voltage_bounds}, and \eqref{eq:gamma_simple_bounds} – \eqref{eq:gamma_sum}.
\subsubsection{TCSC}
TCSC devices adjust the reactance of transmission lines in order to regulate the power flow and enhance system flexibility. Let $\Delta Y_{\text{tcsc}}^t \in \mathbb{C}$ be the admittance variation introduced by a $\text{tcsc} \in \mathcal{D}$ at time $t$, and let it be decomposed as $\Delta Y_{\text{tcsc}}^t = \Delta G_{\text{tcsc}}^t + \jmath \Delta B_{\text{tcsc}}^t$, where $\Delta G_{\text{tcsc}}^t \in \mathbb{R}$ is the conductance variation and $\Delta B_{\text{tcsc}}^t \in \mathbb{R}$ is the susceptance variation.  Let $X_{i,j} \in \mathbb{R}_+$ be the original series reactance of the transmission line $(i,j) \in \mathcal{L}$. Let $\Delta X_{\text{tcsc}}^t \in \mathbb{R}$ be the reactance variation introduced by the TCSC at time $t$, bounded by:
\begin{equation}
    -0.8 X_{i,j} \leq \Delta X_{\text{tcsc}}^t \leq 0.2 X_{i,j} \ .
    \label{eq:tcsc_reactance_bounds}
\end{equation}
 We neglect the conductance of TCSC as in \cite{abdollahi2020optimal} and define the injected admittance as:
\begin{equation}
    Y_{\text{tcsc}}^t = \jmath \Delta B_{\text{tcsc}}^t \ .
    \label{eq:deltaYTCSC}
\end{equation}
Similarly to~\cite{rui2022linear}, we model the behaviour of the susceptance variation $\Delta B_{\text{tcsc}}^t$ caused by the reactance modification $\Delta X_{\text{tcsc}}^t$ using a convex relaxation. Let $s_{\text{tcsc}}^t \in \{0,1\}$ be the binary activation status of a $\text{tcsc} \in \mathcal{D}$ at time $t$, and let $\underline{\Delta X}_{\text{tcsc}}, \overline{\Delta X}_{\text{tcsc}} \in \mathbb{R}$ be the minimum and maximum allowed TCSC reactance variations. The following constraints describe the linearized susceptance behaviour:
\begin{align}
    -\frac{1}{X_{i,j} + \overline{\Delta X}_{\text{tcsc}}} - M(1 - s_{\text{tcsc}}^t)
    &\leq \Delta B_{\text{tcsc}}^t
    \leq -\frac{1}{X_{i,j} + \underline{\Delta X}_{\text{tcsc}}} + M(1 - s_{\text{tcsc}}^t)
    && \label{eq:tcsc_b_limits} \\
    -M s_{\text{tcsc}}^t &\leq \Delta B_{\text{tcsc}}^t \leq M s_{\text{tcsc}}^t \ .
    && \label{eq:tcsc_b_activation}
\end{align}
Using the range values of $\Delta X_{\text{tcsc}}^t$ from~\eqref{eq:tcsc_reactance_bounds} in~\eqref{eq:tcsc_b_limits}, we obtain:

\begin{equation}
    -\frac{1}{1.2 X_{i,j}} - M(1 - s_{\text{tcsc}}^t)
    \leq \Delta B_{\text{tcsc}}^t
    \leq -\frac{1}{0.2 X_{i,j}} + M(1 - s_{\text{tcsc}}^t) \ . 
    \label{eq:tcsc_b_range_numeric}
\end{equation}
Combining together,~\eqref{eq:deltaYTCSC}, \eqref{eq:tcsc_b_activation}, and \eqref{eq:tcsc_b_range_numeric} model TCSC injections.
%
\subsection{Control constraints}
Let $s_{\text{dev}}^t \in \{0,1\}$ be the binary variable used to activate or deactivate $\text{dev} \in \mathcal{D}$, where the type $m \in \{\text{shunt}, \text{stat}, \text{tcsc}\}$ at time $t \in \mathcal{T}$. Let $o_{\text{tap}}^t \in \{0,1\}$ be the indicator variable for a switching action between two consecutive OLTC tap positions. The number of actions allowed at time $t$ is limited to $\overline{a}$, which accounts for both discrete device activations and OLTC tap changes. This constraint is expressed by:
\begin{equation}
    \sum_{\substack{\text{dev} \in \mathcal{D} \setminus \{\text{OLTC}\}}} \left| s_{\text{dev}}^t - s_{\text{dev}}^{t-1} \right|
    + \sum_{\text{tap} \in \{\text{OLTC}\}} o_{\text{tap}}^t
    \leq \overline{a} \ .\label{eq:max_switching_constraint}
\end{equation}
Let $\Delta_{\text{tap}} \in \mathbb{N}$ be the maximum number of OLTC tap steps allowed per time step. The OLTC switching behaviour links the change in discrete tap position with the activation indicator $o_{\text{tap}}^t$. This relation is bounded for $t > 0$ by $\Delta_{\text{tap}}$, as follows:
\begin{equation}
    o_{\text{tap}}^t \leq \left| u_{\text{tap}}^t - u_{\text{tap}}^{t-1} \right| \leq o_{\text{tap}}^t \Delta_{\text{tap}} \ .
    \label{eq:oltc_switch_bounds}
\end{equation}
Because~\eqref{eq:oltc_switch_bounds} is nonconvex due to the presence of a lower bound on the absolute value, we linearize it using an auxiliary variable $\eta_{\text{tap}}^t \in \mathbb{R}_+$ to represent the tap change magnitude. Let $z_{\text{tap}}^t \in \{0,1\}$ be a binary variable used to distinguish between the two cases $u_{\text{tap}}^t \geq u_{\text{tap}}^{t-1}$ and $u_{\text{tap}}^t < u_{\text{tap}}^{t-1}$. The linearized formulation is given by:
\begin{align}
    & \eta_{\text{tap}}^t \geq u_{\text{tap}}^t - u_{\text{tap}}^{t-1} \label{eq:delta_case1} \\
    & \eta_{\text{tap}}^t \geq u_{\text{tap}}^{t-1} - u_{\text{tap}}^t \label{eq:delta_case2} \\
    & \eta_{\text{tap}}^t \leq u_{\text{tap}}^t - u_{\text{tap}}^{t-1} + M (1 - z_{\text{tap}}^t) \label{eq:delta_case1_bigM} \\
    & \eta_{\text{tap}}^t \leq u_{\text{tap}}^{t-1} - u_{\text{tap}}^t + M z_{\text{tap}}^t \label{eq:delta_case2_bigM} \\
    & o_{\text{tap}}^t \leq \eta_{\text{tap}}^t \leq o_{\text{tap}}^t \eta_{\text{tap}} \ . \label{eq:delta_activation_bounds}
\end{align}
In addition, the integer tap position $u_{\text{tap}}^t$ is encoded using the binary vector $\bm{\alpha}_{\text{tap}}^t$ as follows:
\begin{equation}
    u_{\text{tap}}^t = \sum_{n \in \mathcal{K}_{\text{tap}}} n \alpha_{\text{tap},n}^t \ .
    \label{eq:oltc_tap_integer}
\end{equation}
The domains of all control and auxiliary variables are provided by:
\begin{equation}
\begin{aligned}
    & s_{\text{dev}}^t \in \{0,1\}, \quad
    o_{\text{tap}}^t \in \{0,1\}, \quad
    z_{\text{tap}}^t \in \{0,1\},
    \quad u_{\text{tap}}^t \in \mathcal{K}_{\text{tap}}, \quad
    \eta_{\text{tap}}^t \in \mathbb{N_+} \\\
    & \bm{\alpha}_{\text{tap}}^t \in \{0,1\}^K, \quad
    \bm{\alpha}_{\text{shunt}}^t \in \{0,1\}^N .
\end{aligned}
\label{eq:domain_all_controls}
\end{equation}
Finally, the control constraints retained for this problem are given by~\eqref{eq:max_switching_constraint} and  \eqref{eq:delta_case1} – \eqref{eq:domain_all_controls}.

\subsection{Power flow constraints}
For simplicity, we define the active power loss as the sum of directional power flows on each line. Let $p_{i,j}^t \in \mathbb{R}$ be the active power flowing from node $i$ to node $j$ on line $(i,j) \in \mathcal{L}$ at time $t \in \mathcal{T}$, and $p_{j,i}^t \in \mathbb{R}$ the corresponding flow in the reverse direction. Let $P_{i,j,\text{Loss}}^t \in \mathbb{R}_+$ be the active power loss on line $(i,j)$ at time $t$, given by:

\begin{equation}
    P_{i,j,\text{Loss}}^t = p_{i,j}^t + p_{j,i}^t \ .
    \label{eq:line_loss}
\end{equation}
We use the second-order cone relaxation of the optimal power flow formulation~\cite{jabr2006radial}, incorporating the modified admittance introduced by the TCSC device. Let $\mathbf{W}^t \in \mathbb{C}^{|\mathcal{N}| \times |\mathcal{N}|}$ be the Hermitian lifting variable that aims to represent the product $\mathbf{v}^t(\mathbf{v}^t)^\mathrm{H}$ at time $t \in \mathcal{T}$, where $\mathbf{v}$ is the vector collecting all nodal voltage phasors within the network. Let $Y_{i,j} \in \mathbb{C}$ be the admittance of the line $(i,j) \in \mathcal{L}$, and let $\underline{v}_i, \overline{v}_i \in \mathbb{R}_+$ denote the minimum and maximum allowed voltage magnitudes at bus $i \in \mathcal{N}$. The power flow constraints are expressed as follows:
\begin{alignat}{2}
    p_{i,j}^t + \jmath q_{i,j}^t &= (W_{i,i}^t - W_{i,j}^t) Y_{ij}^* 
    &\quad& \forall (i,j) \in \mathcal{L} \setminus \mathcal{L}^{\text{TCSC}},\; t \in \mathcal{T} \label{eq:complex_power_flow} \\
    p_{i,j}^t + \jmath q_{i,j}^t &= (W_{i,i}^t - W_{i,j}^t)(Y_{ij} + \Delta Y_{\text{TCSC}}^t)^*
    &\quad& \forall (i,j) \in \mathcal{L}^{\text{TCSC}},\; t \in \mathcal{T} \label{eq:complex_power_flow_tcsc} \\
    |W_{i,j}^t|^2 &\leq W_{i,i}^t W_{j,j}^t
    &\quad& \label{eq:Wij_sdp} \forall (i,j) \in \mathcal{N},\; t \in \mathcal{T} \\
    \underline{v}_i^2 &\leq W_{i,i}^t \leq \overline{v}_i^2
    &\quad& \label{eq:voltage_magnitude_bounds} \forall i \in \mathcal{N},\; t \in \mathcal{T}\\
    W_{i,i}^t &= U_{\text{tap}}^t 
    &\quad& \forall i \in \mathcal{N}^{\text{OLTC}},\; t \in \mathcal{T} \ . \label{eq:Wi_equals_Utap}
\end{alignat} Constraint~\eqref{eq:complex_power_flow_tcsc} is nonconvex due to the bilinear terms. We apply the reformulation-linearization technique (RLT) inequalities, as described in~\cite{bingane2019tight}, to convexify the bilinear terms $W_{i,i}^t \Delta Y_{\text{tcsc}}^t$ and $W_{i,j}^t \Delta Y_{\text{tcsc}}^t$ using auxiliary variables $f_{\text{tcsc}}^t \in \mathbb{R}$ and $g_{\text{tcsc}}^t \in \mathbb{R}$, respectively. Additionally, from~\eqref{eq:tcsc_b_range_numeric}, we determine that $\Delta Y_{\text{tcsc}}^t$ is bounded between $-\frac{1}{1.2 X_{i,j}}$ and $-\frac{1}{0.2 X_{i,j}}$.
First, we model the convex envelope of $W_{i,i}^t \Delta Y_{\text{tcsc}}^t$ using the auxiliary variable $f_{\text{tcsc}}^t$. The resulting inequalities are:
\begin{align}
    \Delta B_{\text{tcsc}}^t \underline{v}_i^2 - \frac{W_{i,i}^t}{0.2 X_{i,j}} + \frac{\underline{v}_i^2}{0.2 X_{i,j}}
    &\leq f_{\text{tcsc}}^t
    \leq \Delta B_{\text{tcsc}}^t  \underline{v}_i^2 - \frac{W_{i,i}^t}{1.2 X_{i,j}} + \frac{\underline{v}_i^2}{1.2 X_{i,j}}
    \label{eq:f_tcsc_envelope1} \\
    \Delta B_{\text{tcsc}}^t  \overline{v}_i^2 - \frac{W_{i,i}^t}{1.2 X_{i,j}} + \frac{\overline{v}_i^2}{1.2 X_{i,j}}
    &\leq f_{\text{tcsc}}^t
    \leq \Delta B_{\text{tcsc}}^t  \overline{v}_i^2 - \frac{W_{i,i}^t}{0.2 X_{i,j}} + \frac{\overline{v}_i^2}{0.2 X_{i,j}} \ .
    \label{eq:f_tcsc_envelope2}
\end{align}
Then, we model the convex envelope of $W_{i,j}^t \Delta Y_{\text{tcsc}}^t$ using the auxiliary variable $g_{\text{tcsc}}^t$ as follows:
\begin{align}
    \Delta B_{\text{tcsc}}^t  \underline{v}_i^2 - \frac{W_{i,j}^t}{0.2 X_{i,j}} + \frac{\underline{v}_i^2}{0.2 X_{i,j}}
    &\leq g_{\text{tcsc}}^t
    \leq \Delta B_{\text{tcsc}}^t  \underline{v}_i^2 - \frac{W_{i,j}^t}{1.2 X_{i,j}} + \frac{\underline{v}_i^2}{1.2 X_{i,j}}
    \label{eq:g_tcsc_envelope1} \\
    \Delta B_{\text{tcsc}}^t  \overline{v}_i^2 - \frac{W_{i,j}^t}{1.2 X_{i,j}} + \frac{\overline{v}_i^2}{1.2 X_{i,j}}
    &\leq g_{\text{tcsc}}^t
    \leq \Delta B_{\text{tcsc}}^t  \overline{v}_i^2 - \frac{W_{i,j}^t}{0.2 X_{i,j}} + \frac{\overline{v}_i^2}{0.2 X_{i,j}} \ . 
    \label{eq:g_tcsc_envelope2}
\end{align}
Using $f_{\text{tcsc}}^t$ and $g_{\text{tcsc}}^t$, we obtain:
\begin{equation}
    p_{i,j}^t + \jmath q_{i,j}^t = (W_{i,i}^t - W_{i,j}^t) Y_{i,j}^* + \jmath(f_{\text{tcsc}}^t - g_{\text{tcsc}}^t) \ . 
    \label{eq:power_flow_convexified}
\end{equation}
The nonconvex constraint \eqref{eq:complex_power_flow_tcsc} is replaced by the convex relaxation composed of \eqref{eq:f_tcsc_envelope1} – \eqref{eq:power_flow_convexified}.
Let $\overline{S}_{i,j} \in \mathbb{R}_{+}$ be the apparent power flow limit on line $(i,j) \in \mathcal{L}$. The line thermal limit constraint is written as follows:
\begin{equation}
    (p_{i,j}^t)^2 + (q_{i,j}^t)^2 \leq \overline{S}_{i,j}^2 \ .
    \label{eq:thermal_limit}
\end{equation}
Let $p_{\text{dem},i}^t, q_{\text{dem},i}^t \in \mathbb{R}_+$ be the active and reactive power demands, respectively, at bus $i \in \mathcal{N}$ and time $t \in \mathcal{T}$.  
Let $p_{\text{gen},i}^t, q_{\text{gen},i}^t \in \mathbb{R}_+$ be the active and reactive power generations at the same bus and time.  
Let $q_{\text{dev},i}^t \in \mathbb{R}$ represent the total reactive power injection at bus $i$ and time $t$ from $\text{dev} \in \mathcal{D}$ of type $m \in \{\text{shunt}, \text{stat}\}$.  
The nodal power balance is represented below, considering the reactive power injections from shunts and STATCOMs:
\begin{align}
    p_{\text{dem},i}^t - p_{\text{gen},i}^t &= \sum_{(i,j) \in \mathcal{L}} p_{i,j}^t
    \label{eq:active_power_balance} \\
    q_{\text{dem},i}^t - q_{\text{gen},i}^t + \sum_{\text{dev} \in \mathcal{D}} q_{\text{dev},i}^t &= \sum_{(i,j) \in \mathcal{L}} q_{i,j}^t \ .
    \label{eq:reactive_power_balance}
\end{align}
Let $\underline{p}_i, \overline{p}_i \in \mathbb{R}$ and $\underline{q}_i, \overline{q}_i \in \mathbb{R}$ be the minimum and maximum allowable active and reactive power injections at bus $i \in \mathcal{N}$.  
The nodal active and reactive power limits are given as follows:
\begin{align}
    \underline{p}_i \leq p_i^t \leq \overline{p}_i \\
    \underline{q}_i \leq q_i^t \leq \overline{q}_i  &\ . 
    \label{eq:nodal_power_limits}
\end{align}

\subsection{Objective function and full problem}
The main objective of our problem is to minimize the total active power loss while adhering to power flow,
operational, and control constraints. The problem is formulated as an MISOCP as follows:
\begin{equation}
\begin{aligned}
    \min_{\bm{\mathcal{X}}} \quad & P_\text{loss}=\sum_{t \in \mathcal{T}} \sum_{(i,j) \in \mathcal{L}} P_{i,j,\text{Loss}}^t \\
    \text{s.t.} \quad 
    & \eqref{eq:shunt_one_hot}-\eqref{eq:shunt_Mcap},\; \eqref{eq:stat_q_abs_bigM},\; \eqref{eq:tap_binary_one_hot},\; \eqref{eq:tap_squared_voltage_bounds}, \\
    & \eqref{eq:gamma_simple_bounds}-\eqref{eq:gamma_sum},\; \eqref{eq:deltaYTCSC},\; \eqref{eq:tcsc_b_activation}-\eqref{eq:max_switching_constraint}, \\
    & \eqref{eq:delta_case1}-\eqref{eq:complex_power_flow},\; \text{and}~\eqref{eq:Wij_sdp}-\eqref{eq:nodal_power_limits}.
\end{aligned}
\label{eq:opf_objective}
\end{equation}
where
\begin{align*}
\bm{\mathcal{X}} := \Big\{
& p_{i,j}^t,\; q_{i,j}^t,\; P_{i,j,\text{Loss}}^t,\; p_i^t,\; q_i^t,\; p_{\text{dem}}^t,\; p_{\text{gen}}^t,\; q_{\text{dem}}^t,\; q_{\text{gen}}^t,\; q_{\text{dev}}^t,\;
U_{\text{tap}}^t,\; \tilde{U}_{\text{tap}}^t,\; W_{i,i}^t,\; W_{i,j}^t,\; \Delta Y_{\text{TCSC}}^t,\; \Delta X_{\text{TCSC}}^t,\; \Delta  B_{\text{TCSC}}^t,\; \\ & \quad f_{\text{TCSC}}^t,\; g_{\text{TCSC}}^t,\;  s_{\text{dev}}^t,\; o_{\text{tap}}^t,\; \eta_{\text{tap}}^t,\; u_{\text{tap}}^t,\; z_{\text{tap}}^t,\; \bm{\alpha}_{\text{tap}}^t,\; \bm{\gamma}_{\text{tap}}^t,\; \bm{\alpha}_{\text{shunt}}^t
\Big\}.
\end{align*}
Note that other objectives can be considered instead of power loss, such as, minimizing nominal voltage deviation, line overloading, generation costs, and switching costs.
\section{Results and Discussion}
This section presents a numerical evaluation of the proposed MISOCP framework for coordinated FACTS scheduling. The model is applied to three benchmark systems: IEEE 9-bus, IEEE 30-bus, and RTS96. Each case is simulated over an hourly horizon with time-varying load profiles. A switching constraint limits the number of allowable device activations at each time step.

The study assesses: (i) voltage regulation; (ii) active power loss reduction; (iii) MISOCP scalability with network size; and (iv) solver efficiency. Table~\ref{tab:misocp_visual_circles_devices_time_var} summarizes key metrics: number of devices $|\mathcal{D}|$, switching limit $\overline{a}$, total generation $p_{\mathrm{gen}}$, power loss ratio $P_{\mathrm{loss}}\%$, total loss $P_{\mathrm{loss}}$, approximated voltage magnitude average $\mathbb{E}[W_{i,i}]$ and variance $\mathbb{V}[W_{i,i}]$, and total solve time $t_\text{solve}$. Various FACTS configurations are tested across the three systems.
\begin{table*}[h!]
    \centering
        \caption{\centering MISOCP results on IEEE 9-bus, IEEE 30-bus, and RTS-96 systems under different FACTS configurations. The worst and best values per performance indicator and test are shown in red and in green, respectively.}
        \label{tab:misocp_visual_circles_devices_time_var}
    \begin{tabular}{|c|c|c|c|c|c|c|c|c|c|c|}
    \hline

    \hline
    \textbf{System} & \textbf{Scenario} & $|\mathcal{D}|$ & \( \overline{a} \) & \( p_{\mathrm{gen}} \) [MW] & \( P_{\mathrm{loss}}\% \) & \( P_{\mathrm{loss}} \) [MW] & \( \mathbb{E}[W_{i,i}] \) & \( \mathbb{V}[W_{i,i}] \) & \( T \) & \( t_{\text{solve}} \) [s] \\
    \hline

    \hline
    IEEE 9-Bus & \emph{Baseline} & 0 & 0 & \textcolor{red}{1009.51} & \textcolor{red}{3.031} & \textcolor{red}{30.59} & 1.17092 & 0.00536 & 24 & \textcolor{green!60!black}{41.26} \\ \hline
    IEEE 9-Bus & \emph{OLTC} & 1 & 1 & 1009.51 & 3.031 & 30.59 & 1.17023 & 0.00551 & 24 & 93.71 \\ \hline
    IEEE 9-Bus & \emph{STATCOM} & 1 & 1 & 1009.49 & 3.027 & 30.55 & 1.17031 & 0.00551 & 24 & 44.25 \\ \hline
    IEEE 9-Bus & \emph{SHUNT} & 1 & 1 & 1009.51 & 3.030 & 30.59 & \textcolor{red}{1.17284} & \textcolor{green!60!black}{0.00535} & 24 & 46.08 \\ \hline
    IEEE 9-Bus & \emph{TCSC} & 1 & 1 & 1009.46 & 3.021 & 30.49 & \textcolor{green!60!black}{1.16691} & 0.00676 & 24 & 50.87 \\ \hline
    IEEE 9-Bus & \emph{All devices} & 4 & 2 & \textcolor{green!60!black}{1009.44} & \textcolor{green!60!black}{3.016} & \textcolor{green!60!black}{30.45} & 1.17003 & \textcolor{red}{0.00738} & 24 & \textcolor{red}{97.28} \\
    \hline\hline
    IEEE 30-Bus & \emph{Baseline} & 0 & 0 & \textcolor{red}{2613.35} & \textcolor{red}{1.860} & \textcolor{red}{48.60} & \textcolor{green!60!black}{1.15572} & \textcolor{red}{0.00217} & 12 & \textcolor{green!60!black}{137.29} \\ \hline
    IEEE 30-Bus & \emph{OLTC} & 3 & 1 & 2613.35 & 1.860 & 48.60 & 1.15572 & 0.00217 & 12 & 268.68 \\ \hline
    IEEE 30-Bus & \emph{STATCOM} & 3 & 1 & 2609.64 & 1.578 & 41.18 & \textcolor{red}{1.18893} & \textcolor{green!60!black}{0.00027} & 12 & 167.01 \\ \hline
    IEEE 30-Bus & \emph{SHUNT} & 2 & 1 & 2609.91 & 1.797 & 41.72 & 1.17645 & 0.00027 & 12 & \textcolor{red}{285.14} \\ \hline
    IEEE 30-Bus & \emph{TCSC} & 3 & 1 & 2611.76 & 1.739 & 45.42 & 1.17241 & 0.00186 & 12 & 262.27 \\ \hline
    IEEE 30-Bus & \emph{All devices} & 6 & 2 & \textcolor{green!60!black}{2609.39} & \textcolor{green!60!black}{1.559} & \textcolor{green!60!black}{40.67} & 1.18830 & 0.00034 & 12 & 278.8 \\
    \hline\hline
    RTS-96 & \emph{Baseline} & 0 & 0 & \textcolor{red}{36352.97} & \textcolor{red}{1.355} & \textcolor{red}{492.75} & \textcolor{green!60!black}{1.156508} & \textcolor{red}{0.00291} & 12 & \textcolor{green!60!black}{1444.05} \\ \hline

    RTS-96 & \emph{OLTC} & 8 & 2 & 36352.97 & 1.355 & 492.75 & 1.17742 & 0.00032 & 12 & 2076.88 \\ \hline
    RTS-96 & \emph{STATCOM} & 7 & 2 & 36343.65 & 1.305 & 474.12 & 1.17660 & 0.00101 & 12 & 1531.17 \\ \hline
    RTS-96 & \emph{SHUNT} & 3 & 2 & 36346.46 & 1.320 & 479.73 & 1.16379 & 0.00257 & 12 & 1911.76 \\ \hline
    RTS-96 & \emph{TCSC} & 4 & 2 & 36340.13 & 1.285 & 467.08 & 1.18099 & 0.00051 & 12 & 2074.63 \\ \hline
    RTS-96 & \emph{All devices} & 15 & 3 & \textcolor{green!60!black}{36335.68} & \textcolor{green!60!black}{1.261} & \textcolor{green!60!black}{458.18} & \textcolor{red}{1.18658} & \textcolor{green!60!black}{0.00030}& 12 & \textcolor{red}{2080.13} \\ \hline

\hline    
    \end{tabular}
\end{table*}

On the 9-bus system, the \emph{Baseline} active power loss level is 3.031\%. The \emph{OLTC} maintains the same loss level, while \emph{STATCOM}, \emph{SHUNT}, and \emph{TCSC} reduce it down to 3.027\%, 3.030\%, and 3.021\%, respectively. The \emph{All-devices} configuration achieves the lowest loss at 3.016\%. The approximated voltage magnitude average $\mathbb{E}[W_{i,i}]$ decreases from 1.17092 p.u.~(\emph{Baseline}) to 1.17023 p.u.~(\emph{OLTC}) and 1.16691 p.u.~(\emph{TCSC}). The approximated voltage magnitude variance $\mathbb{V}[W_{i,i}]$ is lowest with \emph{SHUNT} (0.00535) and highest with \emph{All devices} (0.00738). The solve time increases from 41.26\,seconds (\emph{Baseline}) to 97.28\,seconds (\emph{All devices}) representing a 2.36 time slowdown due to added complexity. \emph{OLTC} and \emph{TCSC} require 93.71\,seconds and 50.87\,seconds, respectively.

On the 30-bus system, loss is reduced from 1.860\% (\emph{Baseline}) to 1.797\% (\emph{SHUNT}), 1.739\% (\emph{TCSC}), and 1.578\% (\emph{STATCOM}), reaching 1.559\% with \emph{All devices}. The approximated voltage magnitude average $\mathbb{E}[W_{i,i}]$ rises with \emph{STATCOM} (1.18893 p.u.) and \emph{All devices} (1.18830 p.u.), and $\mathbb{V}[W_{i,i}]$ reaches a minimum of 0.00027 with \emph{STATCOM} and \emph{SHUNT}. The solve time grows from 137.29\,seconds (\emph{Baseline}) to 278.8\,seconds (\emph{All devices}), with \emph{SHUNT} notably requiring the highest computation time (285.14\,seconds).

For RTS-96, the \emph{Baseline} loss of 1.355\% decreases to 1.305\% (\emph{STATCOM}), 1.320\% (Shunt), 1.285\% (\emph{TCSC}), and 1.261\% (\emph{All devices}). The approximated voltage magnitude average $\mathbb{E}[W_{i,i}]$ increases from 1.156508 p.u.~to 1.18658 p.u.~under \emph{All devices}, and $\mathbb{V}[W_{i,i}]$ is minimized with \emph{All devices} (0.00030) and maximized with \emph{Baseline} (0.00291). The solve time rises from 1444.05\,seconds (\emph{Baseline}) to 2080.13\,seconds (\emph{All devices}), with \emph{STATCOM}, \emph{SHUNT}, \emph{TCSC} and \emph{OLTC} ranging between 1531\,seconds and 2077\,seconds.

\begin{figure}[h!]
    \centering
    \includegraphics[width=0.9\textwidth]{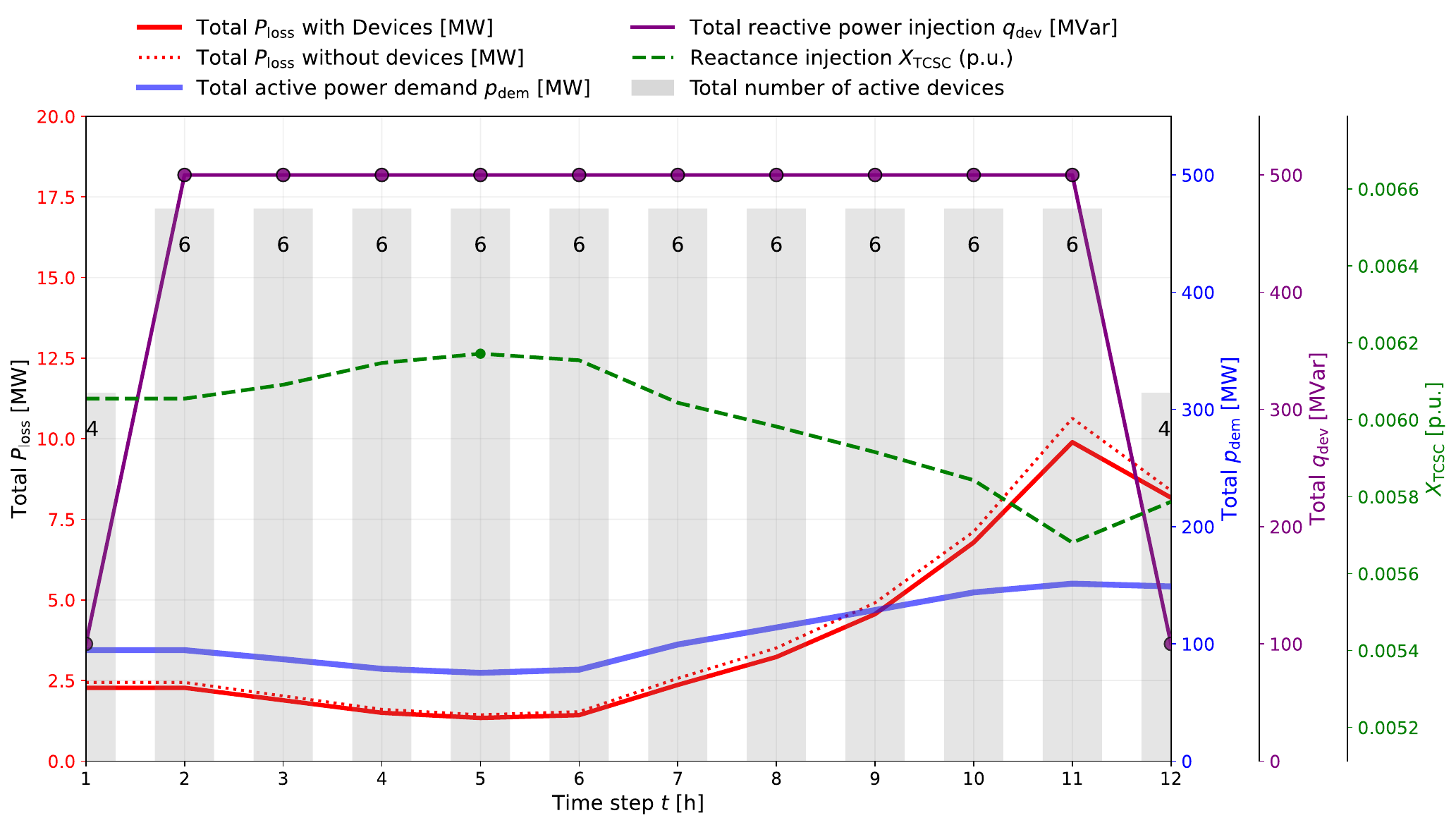}
    \caption{\centering
    Impact of device reactive injections, reactance injections, switching actions, and demand variations on total active power losses over the optimization horizon. Controlled and baseline scenarios are compared.
    }
    \label{fig:results}
\end{figure}

Figure~\ref{fig:results} shows the time evolution of power losses, demand, reactive power injection, TCSC reactance, and device count over a 12-hour horizon on the IEEE 30-bus system under the all-device configuration. The number of active devices stays high at 6 from hour 1 to 10, indicating sustained FACTS involvement. Total reactive power injection $q_{\mathrm{dev}}$ is equal to 500~MVAr at most time steps, while TCSC reactance $X_{\mathrm{tcsc}}$ ranges from 0.0056 to 0.00617~p.u., showing time-varying compensation. Compared to the baseline, the coordinated case reduces active power losses at all time steps, with the largest drop at hour 11—from 10.6~MW to 9.3~MW—an improvement of about 12\%. These results confirm the effectiveness of coordinated FACTS control in adapting to demand variations and minimizing losses. We set $\underline{v}_i$ and $\overline{v}_i$ are equal to 0.9 p.u. and 1.1 p.u. respectively, in the numerical simulations.

The coordination plans generated by our MISOC formulation rely on assumptions that may limit the solution. First, FACTS device placement and parameters are predefined rather than jointly optimized, potentially leading to suboptimal configurations. Second, while voltage limits are enforced via second-order cone relaxation on \( \mathbf{W}^t \), this does not ensure feasibility under the original nonconvex AC power flow~\cite{lesage2020soc}. As a convex results, our relaxations may require reinforcement through heuristic tuning or nonlinear programming solvers to recover AC-feasibility \cite{taylor2013convex}.~The exactness of the relaxation can be checked, for example
with power flow or rank checks~\cite{bahrami2017sdp}. The current examples use simulated data, but real-world deployment will need to handle uncertainty from the environment noises and forecast errors. Finally, to ensure scalability, decomposition methods could be leveraged. These are topics for future work.
\section{Conclusion}
In this work, we propose a mixed-integer second-order cone program (MISOCP) for the multi-period coordination of FACTS devices in power transmission systems. The model integrates OLTCs, STATCOMs, shunt reactors, and TCSCs within a convexified power flow formulation, using binary expansions, disjunctive constraints, reformulation-linearization technique, and a conic relaxation. The resulting MISOCP is compatible with off-the-shelf solvers and enables tractable active power loss minimization over time. The proposed model is evaluated on the IEEE 9-bus, 30-bus, and RTS96 systems. Numerical results illustrates that individual FACTS device configurations (STATCOM, shunt, and TCSC) each reduce losses compared to the baseline, while the full coordination of all device consistently achieves the best performance across all networks. In particular, the all-devices setup yields lower active power losses, though at the cost of increased computation time. This highlights the trade-off between performance gains and computational burden in a coordinated control scheme. Our formulation relies on convex relaxations, and although conic approximations offer computational tractability, these relaxations may yield solutions that are infeasible under the original nonconvex AC model. Future work will focus on evaluating learning-based control approaches, such as multi-agent proximal policy optimization (MAPPO), for coordinating FACTS devices in dynamic grid environments. In addition, decomposition strategies will be explored to enhance scalability, computational efficiency, and real-time feasibility for large-scale transmission networks.

\section*{Acknowledgements}
This work is supported by the Natural Sciences and Engineering Research Council (NSERC) of Canada Alliance-Mitacs Accelerate grant ALLRP~571311-21 ("Optimization of future energy systems") in collaboration with Hydro-Québec.

\newpage
{\fontsize{9pt}{11pt}\selectfont  

\bibliography{ref.bib}

\begin{thebibliography}{10}

\bibitem{kundur1994power}
Kundur P.
\newblock Power System Stability and Control.
\newblock McGraw-Hill Education; 1994.

\bibitem{hatziargyriou2020advanced}
Hatziargyriou N, editor.
\newblock Electricity Supply Systems of the Future.
\newblock CIGRÉ Green Books, Springer; 2020.

\bibitem{hingorani2000facts}
Hingorani NG, Gyugyi L.
\newblock Understanding FACTS: Concepts and Technology of Flexible AC Transmission Systems.
\newblock IEEE Press; 2000.

\bibitem{zhang2012facts}
Zhang XP, Rehtanz C, Pal B.
\newblock Flexible AC Transmission Systems: Modelling and Control.
\newblock Heidelberg, Germany: Springer; 2012.

\bibitem{wood2013power}
Wood AJ, Wollenberg BF, Sheble GB.
\newblock Power Generation, Operation, and Control.
\newblock John Wiley \& Sons; 2013.

\bibitem{zhu2015optimization}
Zhu J.
\newblock Optimization of Power System Operation.
\newblock John Wiley \& Sons; 2015.

\bibitem{jabr2006radial}
Jabr RA.
\newblock Radial Distribution Load Flow Using Conic Programming.
\newblock IEEE Transactions on Power Systems. 2006;21(3):1458-9.

\bibitem{aigner2023globalopf}
Aigner KM, Burlacu R, Liers F, Martin A.
\newblock Solving AC Optimal Power Flow With Discrete Decisions to Global Optimality.
\newblock INFORMS Journal on Computing. 2023 Mar;35(2):458-74.

\bibitem{kayacik2021misocp}
Kayacık SE, Kocuk B.
\newblock An MISOCP-Based Solution Approach to the Reactive Optimal Power Flow Problem.
\newblock IEEE Transactions on Power Systems. 2021 Jan;36(1):529-38.

\bibitem{han2023bumpless}
Han T, Low SH, Weng Y.
\newblock Bumpless Topology Control in Power Networks: A Convex Optimization Approach.
\newblock IEEE Transactions on Power Systems. 2023.

\bibitem{ibrahim2022alternating}
Ibrahim T, Rubira TTD, Rosso AD, Patel M, Guggilam S, Mohamed AA.
\newblock Alternating Optimization Approach for Voltage-Secure Multi-Period Optimal Reactive Power Dispatch.
\newblock IEEE Transactions on Power Systems. 2022;37(5):3805-16.

\bibitem{dey2022power}
Dey S, Deka N, Hazarika D.
\newblock Power System Planning for Reduction in System Losses Using STATCOM and PSO Technique.
\newblock Journal of The Institution of Engineers (India): Series B. 2022 Dec;103(6):1269-81.

\bibitem{mumtahina2023literature}
Mumtahina U, Alahakoon S, Wolfs P.
\newblock A Literature Review on the Optimal Placement of Static Synchronous Compensator (STATCOM) in Distribution Networks.
\newblock Energies. 2023;16(17):6122.

\bibitem{aljumah2024tcsc}
Aljumah AS, Alqahtani MH, Shaheen AM, Elsayed AM.
\newblock Enhancing Power System Performance via TCSC Technology Allocation With Enhanced Gradient-Based Optimization Algorithm.
\newblock IEEE Access. 2024 Jul;12:97806-19.

\bibitem{tarafdar2025factsreview}
Hagh MT, Borhany MAJ, Taghizad-Tavana K, Oskouei MZ.
\newblock A Comprehensive Review of Flexible Alternating Current Transmission System (FACTS): Topologies, Applications, Optimal Placement, and Innovative Models.
\newblock Heliyon. 2025 Jan;11(1):e41001.

\bibitem{bingane2019tight}
Bingane C, Anjos MF, Digabel SL.
\newblock Tight-and-Cheap Conic Relaxation for the Optimal Reactive Power Dispatch Problem.
\newblock IEEE Transactions on Power Systems. 2019 Nov;34(6):4684-93.

\bibitem{sulaiman2022optimal}
Sulaiman MH, Mustaffa Z.
\newblock Optimal Placement and Sizing of FACTS Devices for Optimal Power Flow Using Metaheuristic Optimizers.
\newblock Results in Control and Optimization. 2022;8:100145.

\bibitem{khan2021novel}
Khan NH, Haidar AMA, Hannan MA, Ker PJ.
\newblock A Novel Modified Lightning Attachment Procedure Optimization Technique for Optimal Allocation of the FACTS Devices in Power Systems.
\newblock IEEE Access. 2021;9:47976-97.

\bibitem{iria2019optimal}
Iria J, Heleno M, Cardoso G.
\newblock Optimal Sizing and Placement of Energy Storage Systems and Onload Tap Changer Transformers in Distribution Networks.
\newblock Applied Energy. 2019;250:1147-57.

\bibitem{abdollahi2020optimal}
Abdollahi A, Abapour M, Ghasemi A.
\newblock Optimal Power Flow Incorporating FACTS Devices and Stochastic Wind Power Generation Using Krill Herd Algorithm.
\newblock Electronics. 2020;9(6):1043.

\bibitem{rui2022linear}
Rui X, Sahraei-Ardakani M, Nudell TR.
\newblock Linear Modelling of Series FACTS Devices in Power System Operation Models.
\newblock IET Generation, Transmission \& Distribution. 2022 Apr;16(6):1047-63.

\bibitem{lesage2020soc}
Lesage-Landry A, Taylor JA.
\newblock A Second-Order Cone Model Of Transmission Planning with Alternating and Direct Current Lines.
\newblock European Journal of Operational Research. 2020;281(1):174-85.

\bibitem{taylor2013convex}
Taylor JA, Hover FS.
\newblock Conic AC Transmission System Planning.
\newblock IEEE Transactions on Power Systems. 2013 May;28(2):952-9.

\bibitem{bahrami2017sdp}
Bahrami S, Therrien F, Wong VWS, Jatskevich J.
\newblock Semidefinite Relaxation of Optimal Power Flow for AC--DC Grids.
\newblock IEEE Transactions on Power Systems. 2017 Jan;32(1):289-304.

\end{thebibliography}
}

\bibliographystyle{vancouver}

\end{document}